\documentclass{PoS}
\usepackage{amssymb}
\usepackage{bm}
\usepackage{latexsym}

\newcommand{\possy}[2]{\href{http://pos.sissa.it/cgi-bin/reader/contribution.cgi?id=#1}{\tt #2}}

\newcommand{\case}[2]{\ensuremath{{\textstyle\frac{#1}{#2}}}}
\newcommand{\half}{\ensuremath{{\textstyle\frac{1}{2}}}}

\title{$\bm{B^0_q}$-$\bm{\bar{B}^0_q}$ Mixing and Matching with 
\hfill{\sf\small FERMILAB-CONF-09-599-T} \\
Fermilab Heavy Quarks}

\ShortTitle{$B^0_q$-$\bar{B}^0_q$ Mixing and Matching}

\author{R. Todd Evans \\
        Fakult\"at f\"ur Physik,
        Universit\"at Regensburg, Regensburg, Germany}

\author{Elvira G\'amiz,\thanks{Present address Theoretical Physics Department,
		Fermi National Accelerator Laboratory$^\ddagger$}~
        Aida X. El-Khadra \\
        Physics Department,
        University of Illinois, Urbana, Illinois, USA}

\author{\speaker{Andreas S. Kronfeld} \\
		Theoretical Physics Department,
		Fermi National Accelerator Laboratory,\hspace*{-0.4em}
        \thanks{Operated by Fermi Research Alliance, LLC, under Contract
		No.~DE-AC02-07CH11359 with the United States Department of Energy.}~
		Batavia, Illinois, USA \\
		E-mail: \email{ask@fnal.gov}}

\author{Fermilab Lattice and MILC Collaborations}

\abstract{We discuss the matching procedure for heavy-light 4-quark 
operators using the Fermilab method for heavy quarks and staggered 
fermions for light quarks.
These ingredients enable us to construct the continuum-limit 
operator needed to determine the oscillation frequency of neutral 
$B$~mesons.
The matching is then carried out at the one-loop level.
We also present an updated preliminary result for 
the SU(3)-breaking ratio~$\xi$,
based on calculations using the MILC Collaboration's ensembles of 
lattice gauge fields.}

\FullConference{The XXVII International Symposium on Lattice Field 
Theory---LAT2009\\ July 26--31, 2009\\ Peking University, Beijing, China}

\begin{document}

\section{Introduction}
\label{sec:intro}

All neutral mesons---$K^0$, $B^0$, $B_s$, $D^0$---have been observed 
to oscillate from particle to antiparticle.
The oscillation frequency $\Delta M$ tests the Standard Model's 
pattern of flavor violation.
The phenomenology is especially simple for neutral $B$ mesons 
(normal and strange), because the flavor-changing dynamics play out 
predominantly at distances much shorter than the scale of QCD.
In the case of the $B$ mesons, the width difference $\Delta\Gamma$ 
of the two propagating eigenstates also arises predominantly at 
short distances.
It is especially intriguing (at least for now), because measurements 
of $\Delta\Gamma_s$ and the $CP$ phase~$\phi_s$ of the $B_s$ are in 
imperfect agreement with the Standard 
Model~\cite{Lenz:2006hd,Bona:2008jn}.

Neutral $B$ mixing stems from $\Delta B=2$ flavor-changing transitions.
In the Standard Model these arise first at the one-loop level, so 
non-Standard contributions are conceivably of comparable size.
The observables are then (approximately) $\Delta M = 2|M_{12}|$, 
$\Delta\Gamma = 2|\Gamma_{12}|\cos\phi$, and
$\phi=\arg\left(-M_{12}/\Gamma_{12}\right)$,
where $M_{12}$ and $\Gamma_{12}$ are the off-diagonal elements of the 
mass and width matrices of the two-state systems:
\begin{eqnarray}
    M_{12} & = & \frac{G_F^2}{8\pi^2}\frac{M_W^2}{M^2_{B_q}}
        (V^*_{tq}V_{tb})^2S_0(m_t^2/M_W^2)\eta_b(\mu)
        \langle B|\bar{q}_L\gamma_\mu b\bar{q}_L\gamma^\mu b|\bar{B}\rangle
        + \mbox{BSM}, 
    \label{eq:mass} \\
    \Gamma_{12} & = &  -\frac{G_F^2m_b^2}{6\pi M_{B_q}}\left[
        G(V,\mu)\langle B|\bar{q}_L\gamma_\mu 
        b\bar{q}_L\gamma^\mu b|\bar{B}\rangle +
        G_S(V,\mu)\langle B|\bar{q}_Lb\bar{q}_Lb|\bar{B}\rangle \right]
        + \mbox{BSM},
    \label{eq:width}
\end{eqnarray}
where $V$ is the CKM matrix, 
and $S_0$, $\eta_b$, $G$, and $G_S$ are short-distance effects, 
computed in electroweak and QCD perturbation theory.
Contributions beyond the Standard Model (``BSM'') are not written out 
explicitly.
Because of the $V-A$ structure of the electroweak interaction, only the 
left-handed (light) quark field $\bar{q}_L = \bar{q}\half(1+\gamma_5)$
appears.

The remainder of this paper is organized as follows.
Section~\ref{sec:matching} constructs lattice operators with staggered 
light quarks and Fermilab heavy quarks, 
corresponding to the 4-quark operators in Eqs.~(\ref{eq:mass}) 
and~(\ref{eq:width}).
(The construction suffices for any light quark with chiral symmetry 
and heavy quark with heavy-quark symmetry.)
We give a status report of our numerical results in 
Sec.~\ref{sec:mixing}.
Section~\ref{sec:future} summarizes and presents some of our plans for 
the future.

\section{Short-Distance Matching}
\label{sec:matching}

To compute the hadronic matrix elements in Eqs.~(\ref {eq:mass}) 
and~(\ref {eq:width}), one has to derive an expression in lattice 
gauge theory that approximates well
$\bar{q}_L\gamma_\mu b\bar{q}_L\gamma^\mu b$ and
$\bar{q}_Lb\bar{q}_Lb$.
The lattice operators can then be computed, and the numerical and other 
uncertainties estimated, to determine $M_{12}$ and $\Gamma_{12}$.
Similar operators appear BSM, for which the following derivation serves 
as a template.

For the light valence quark we take naive asqtad propagators
\begin{equation}
    \langle\Upsilon(x)\bar{\Upsilon}(y)\rangle_U =
        \Omega(x)\Omega^{-1}(y)\langle\chi(x)\bar{\chi}(y)\rangle_U,
    \label{eq:naive}
\end{equation}
where $\chi$ is the one-component staggered fermion field;
$\Upsilon$ is a 4-component naive field, and $\langle\cdots\rangle_U$ 
denotes the fermion average in a fixed gauge field~$U$.
For the heavy quark we use
\begin{equation}
    \Psi = [1 + d_1(m_0a)\bm{\gamma}\cdot\bm{D}]\psi,
    \label{eq:rot}
\end{equation}
where $\psi$ is the fermion field appearing in the Fermilab 
action~\cite{ElKhadra:1996mp} or an improved action with the same design 
features~\cite{Oktay:2008ex}.
\pagebreak

We aim to construct lattice operators $Q$ and $Q_S$ such that
\begin{eqnarray}
    Q   & \doteq & \bar{q}_L\gamma_\mu b\bar{q}_L\gamma^\mu b + 
        {\rm O}(a^2),
    \label{eq:Qlat} \\
    Q_S & \doteq & \bar{q}_Lb\bar{q}_Lb + {\rm O}(a^2),
    \label{eq:QSlat}
\end{eqnarray}
where $\doteq$ means ``has the same matrix elements as.''
Here the ${\rm O}(a^2)$ term depends on $m_ba$.
As long as one retains small corrections to heavy-quark symmetry, 
it remains bounded even as $m_ba\to\infty$;
as long as certain Dirac off-diagonal improvements are consistently 
introduced~\cite{ElKhadra:1996mp,Oktay:2008ex}, they vanish as $a\to 0$.
These two elements are the essence of the Fermilab method.

Our construction starts with the lattice operators
$\bar{\Upsilon}_L\gamma_\mu \Psi\bar{\Upsilon}_L\gamma^\mu \Psi$ and 
$\bar{\Upsilon}_L\Psi\bar{\Upsilon}_L\Psi$.
According to the HQET theory of cutoff 
effects~\cite{Kronfeld:2000ck,Harada:2001fi,Harada:2001fj},
these lattice operators can be described by
\begin{eqnarray}
    \bar{\Upsilon}_L\gamma_\mu \Psi\bar{\Upsilon}_L\gamma^\mu \Psi & \doteq &
        2C^{\rm lat}\bar{q}_L\gamma_\mu h^{(+)}\bar{q}_L\gamma^\mu h^{(-)} +
        2\delta C^{\rm lat}\bar{q}_L h^{(+)}\bar{q}_L h^{(-)} +
        \sum_{i=1}^5 B^{\rm lat}_i\mathcal{Q}_i + \cdots,
    \label{eq:hqetVVlat} \\
    \bar{\Upsilon}_L\Psi\bar{\Upsilon}_L\Psi & \doteq &
        2\delta C^{\rm lat}_S\bar{q}_L\gamma_\mu h^{(+)}\bar{q}_L\gamma^\mu h^{(-)} +
        2C^{\rm lat}_S\bar{q}_L h^{(+)}\bar{q}_L h^{(-)} +
        \sum_{i=1}^5 B^{\rm lat}_{Si}\mathcal{Q}_i + \cdots,
    \label{eq:hqetSSlat}
\end{eqnarray}
where $h^{(\pm)}$ are the heavy-quark fields of the heavy-quark 
effective theory (HQET), 
satisfying $h^{(\pm)}=\half(1\pm\gamma_4)h^{(\pm)}$.
The sums are over five dimension-7, $\Delta B=2$, four-quark 
operators, similar to those written out, but with an extra derivative.
The series continues with operators of dimension 8 and higher.
On the right-hand side of 
Eqs.~(\ref{eq:hqetVVlat}) and (\ref{eq:hqetSSlat}) 
the operators are to be understood with some continuum regulator and 
renormalization scheme.
Discretization effects are lumped into the short-distance coefficients 
$C^{\rm lat}_{(S)}$, $\delta C^{\rm lat}_{(S)}$, 
and $B^{\rm lat}_{(S)i}$, which depend on the couplings of the lattice 
action, as well as the lattice spacing~$a$ and the (renormalized) gauge 
coupling and quark masses.

The next step is to note that the target operators have a completely 
parallel description in HQET, namely
\begin{eqnarray}
    \bar{q}_L\gamma_\mu b\bar{q}_L\gamma^\mu b & \doteq &
        2C\bar{q}_L\gamma_\mu h^{(+)}\bar{q}_L\gamma^\mu h^{(-)} +
        2\delta C\bar{q}_L h^{(+)}\bar{q}_L h^{(-)} +
        \sum_{i=1}^5 B_i\mathcal{Q}_i + \cdots,
    \label{eq:hqetVV} \\
    \bar{q}_Lb\bar{q}_Lb & \doteq &
        2\delta C_S\bar{q}_L\gamma_\mu h^{(+)}\bar{q}_L\gamma^\mu h^{(-)} +
        2C_S\bar{q}_L h^{(+)}\bar{q}_L h^{(-)} +
        \sum_{i=1}^5 B_{Si}\mathcal{Q}_i + \cdots,
    \label{eq:hqetSS}
\end{eqnarray}
where the (continuum HQET) operators on the right-hand sides of
Eqs.~(\ref{eq:hqetVV}) and (\ref{eq:hqetSS}) 
are precisely the same as those on the right-hand sides of
Eqs.~(\ref{eq:hqetVVlat}) and (\ref{eq:hqetSSlat}).
The coefficients differ, however, because the lattice does not appear 
on the left-hand side of Eqs.~(\ref{eq:hqetVV}) and (\ref{eq:hqetSS}).

With Eqs.~(\ref{eq:hqetVVlat})--(\ref{eq:hqetSS}) the desired 
construction of $Q$ and $Q_S$ is immediate: 
\begin{eqnarray}
    Q   & = & Z\bar{\Upsilon}_L\gamma_\mu\Psi\bar{\Upsilon}_L\gamma^\mu\Psi
        + \delta Z\bar{\Upsilon}_L\Psi\bar{\Upsilon}_L\Psi 
        + \sum_i b_i Q_i, 
    \label{eq:Q} \\
    Q_S & = & Z_S\bar{\Upsilon}_L\Psi\bar{\Upsilon}_L\Psi 
        + \delta Z_S\bar{\Upsilon}_L\gamma_\mu\Psi\bar{\Upsilon}_L\gamma^\mu\Psi
        + \sum_i b_{Si} Q_i,
    \label{eq:QS}
\end{eqnarray}
where the $Q_i$ are lattice discretizations of the $\mathcal{Q}_i$, 
such that $Q_i\doteq C^{\rm lat}_{ij}\mathcal{Q}_j+\mbox{dimension~8}$.
Simple algebra then shows that if
\begin{eqnarray}
    Z & = & \left[CC_S^{\rm lat} - \delta C\delta C_S^{\rm lat}
        \right]/\left[C^{\rm lat}C_S^{\rm lat} - 
            \delta C^{\rm lat} \delta C_S^{\rm lat}\right],
    \label{eq:Z}  \\
    \delta Z & = & \left[\delta C - Z\,\delta C^{\rm lat}\right]/
        C_S^{\rm lat},
    \label{eq:dZ} \\
    b_i & = & 
        \left[B_j - Z\,B_j^{\rm lat} - \delta Z\,B_{Sj}^{\rm lat}\right]
        {C^{\rm lat}}_{ji}^{-1},
    \label{eq:bi}
\end{eqnarray}
then Eq.~(\ref{eq:Qlat}) is satisfied.
Similar expressions exist for $Z_S$, $\delta Z_S$, and $b_{Si}$, such 
that Eq.~(\ref{eq:QSlat}) is satisfied.
From the structure of Eqs.~(\ref{eq:Z})--(\ref{eq:bi}) it is clear that 
the regulator and renormalization scheme dependence of the HQET drops 
out of $Z_{(S)}$, $\delta Z_{(S)}$, and $b_{(S)i}$.

Let us close this section with a few remarks.
The enumeration of the operators $\mathcal{Q}_i$, and further operators 
of dimension 8, is an easy extension of Ref.~\cite{Harada:2001fi}.
In perturbation theory $C_{(S)}$ ($\delta C_{(S)}$ and the $B_i$) 
start at tree (one-loop) level, but they could also be determined 
nonperturbatively, adapting schemes such as that of 
Ref.~\cite{Lin:2006ur}.
Because of the way Fermilab lattice actions are 
constructed~\cite{ElKhadra:1996mp,Oktay:2008ex},
starting with Wilson fermions, one has $\lim_{a\to0}C^{\rm lat}=C$, 
\emph{etc.}, without fine tuning.
(In lattice NRQCD this is possible only with fine tuning.)
Although our derivation hinges on the HQET description of cutoff 
effects, one could also (for $m_ba\ll1$) use the Symanzik theory;
the results for $Z_{(S)}$, $\delta Z_{(S)}$, and $b_{(S)i}$ would be 
the same.

We have embarked on a one-loop calculation of $Z_{(S)}$ 
and~$\delta Z_{(S)}$.
At present they are being checked.
As with currents~\cite{Harada:2001fi,Harada:2001fj}, it may prove 
prudent to write
\begin{equation}
    Z_{(S)} = Z_{V_{bb}}Z_{V_{qq}}\rho_{(S)},
    \label{eq:rho}
\end{equation}
where $Z_{V_{bb}}$ and $Z_{V_{qq}}$ are nonperturbatively determined 
matching factors for the vector current.
The remaining factor $\rho_{(S)}$ could have a tamer perturbative 
expansion, because of cancellation among diagrams.
We do not expect the cancellation to be as good as in the case of 
currents, because 4-quark operators have new diagrams in which a gluon 
is exchanged from one bilinear to the other.

With the rotation of Eq.~(\ref{eq:rot}), the $b_{(S)i}$ in 
Eqs.~(\ref{eq:Q}) and~(\ref{eq:QS}) are of order $\alpha_s$ 
and are not available.
The calculations of the 4-quark operator matrix elements described 
below thus have discretization errors of the form 
\begin{eqnarray}
    \frac{B_{(S)i}\langle\mathcal{Q}_i\rangle}{%
    \langle\bar{q}_L\gamma_\mu b\bar{q}_L\gamma^\mu b\rangle} & \sim &
        a\Lambda\frac{\alpha_s}{2(1+m_0a)} , \\
    \mbox{dim~8~ops} & \sim & a^2\Lambda^2f(m_0a),
\end{eqnarray}
where the mass dependence of the $B_{(S)i}$ is an Ansatz with the 
correct asymptotic behavior as $m_0\to\infty$ and as $m_0a\to0$
for the Fermilab action.
The functions $f(m_0a)$ multiplying the ${\rm O}(a^2)$ discretization 
effects are known for the Fermilab 
action~\cite{Harada:2001fi,Bailey:2008wp}.

\section{Long-Distance Matrix Elements}
\label{sec:mixing}

To compute the matrix elements we use a data-object called the 
open-meson propagator~\cite{Evans:2006zz}.
Valence quark propagators are started at an origin $(\bm{x}_0,t_0)$, 
where the 4-quark operator sits, out to all $(\bm{x},t)$.
Since, for this problem, we are interested only in zero-momentum
pseudoscalars, at each $t$ the Dirac indices are contracted with 
$\gamma_5$, and this contraction is summed over all $\bm{x}$.
On the other hand, $M_{12}$ and $\Gamma_{12}$ require two (several) 
Dirac structures in (beyond) the Standard Model.
Therefore we leave the Dirac and color indices free at 
$(\bm{x}_0,t_0)$, writing out one $12\times12\times N_4$ data-object
per configuration, where $N_4$ is the total number of time slices.
Three-point functions are formed by contracting open-meson propagators 
at times $t_i$ and $t_f$ with the Dirac structure of each 4-quark operator.
Two-point functions from $t_0$ to $t$ are used to normalize the matrix 
elements and to provide a cross-check with our separate calculations of 
$B$-meson decay constants~\cite{Bernard:2009wr}.

Our calculations are carried out on several ensembles of lattice gauge 
fields with a realistic sea of 2+1 flavors, made available by the MILC 
Collaboration~\cite{Bernard:2001av,Aubin:2004wf}. 
The ensembles used here are listed in Table~\ref{tbl:milc} together 
with the valence quark masses.
\begin{table}[bp]
    \centering
    \begin{tabular}{ccccl}
        \hline\hline
        $a$ (fm) & Lattice & $N_{\rm confs}$ & Sea $(am_l,am_h)$ & 
        \multicolumn{1}{c}{Valence $am_q$} \\
        \hline
        0.12       & $24^3\times64$ & 529 & $(0.005,0.05)$ & 
            0.005, 0.007, 0.01, 0.02, 0.03, 0.0415 \\
        ``coarse'' & $20^3\times64$ & 833 & $(0.007,0.05)$ & 
            0.005, 0.007, 0.01, 0.02, 0.03, 0.0415 \\
                   & $20^3\times64$ & 592 &  $(0.01,0.05)$ & 
            0.005, 0.007, 0.01, 0.02, 0.03, 0.0415 \\
                   & $20^3\times64$ & 460 &  $(0.02,0.05)$ & 
            0.005, 0.007, 0.01, 0.02, 0.03, 0.0415 \\
        \hline
        0.09       & $28^3\times96$ & 557 & $(0.0062,0.031)$ & 
            0.0031, 0.0044, 0.062, 0.0124, 0.0272, 0.031 \\
        ``fine''   & $28^3\times96$ & 534 & $(0.0124,0.031)$ & 
            0.0031, 0.0042, 0.062, 0.0124, 0.0272, 0.031 \\
        \hline\hline
    \end{tabular}
    \caption[tbl:milc]{Input parameters for the numerical 
    calculations.
    The lattice spacings listed are approximate mnemonics.
    The heavier sea mass $m_h$ is close to the strange mass, which 
    then is subject to retuning \emph{a~posteriori}, yielding the last
    value of $am_q$ for the coarse ensembles.}
    \label{tbl:milc}
\end{table}
The sea quarks are simulated with the asqtad action for staggered 
quarks, and with the fourth-root procedure to reduce the number of 
species from 4 to~1.

To discuss the analysis, it is helpful to introduce some notation.
The four-quark matrix elements are written
\begin{equation}
    \langle B^0_q|\bar{\Upsilon}_L\gamma_\mu\Psi
            \bar{\Upsilon}_L\gamma^\mu\Psi|\bar{B}^0_q\rangle =
        \case{2}{3}M_{B_q}\beta_q^2,
\end{equation}
where the quantity $\beta_q$ is well-behaved in the heavy-quark limit.
We extract $\beta_s$ and $\beta_d$ from 2- and 3-point functions.
With staggered valence quarks these correlators have contributions 
from wrong-parity states with time dependence $(-1)^{t/a}$.
We are careful to disentangle these states.
To isolate the ground state we use Bayesian fits, varying the number of 
states.

We then carry out a partially-quenched (i.e., $m_q$ and $m_l$ varying
independently) chiral extrapolation of $\beta_q/\beta_s$ to obtain 
$\beta_d/\beta_s$, using rooted staggered chiral perturbation theory 
for $\beta_q$~\cite{BLV:2006,Evans:2008}.
With more valence masses than sea masses, the effects of partial 
quenching constrain the parameters of $\chi$PT more stringently than 
would unitary ($m_q=m_l$) data alone.
Fitting the ratio  $\beta_d/\beta_s$ yields smaller statistical errors 
than fitting $r_1^{3/2}\beta_q$ directly.
We also carry out a chiral extrapolation of $r_1^{3/2}\beta_s$, which 
is mild, because it depends only on the sea masses $(am_l,am_h)$.

\begin{figure}[bp]
    (a)\includegraphics[width=0.45\textwidth]{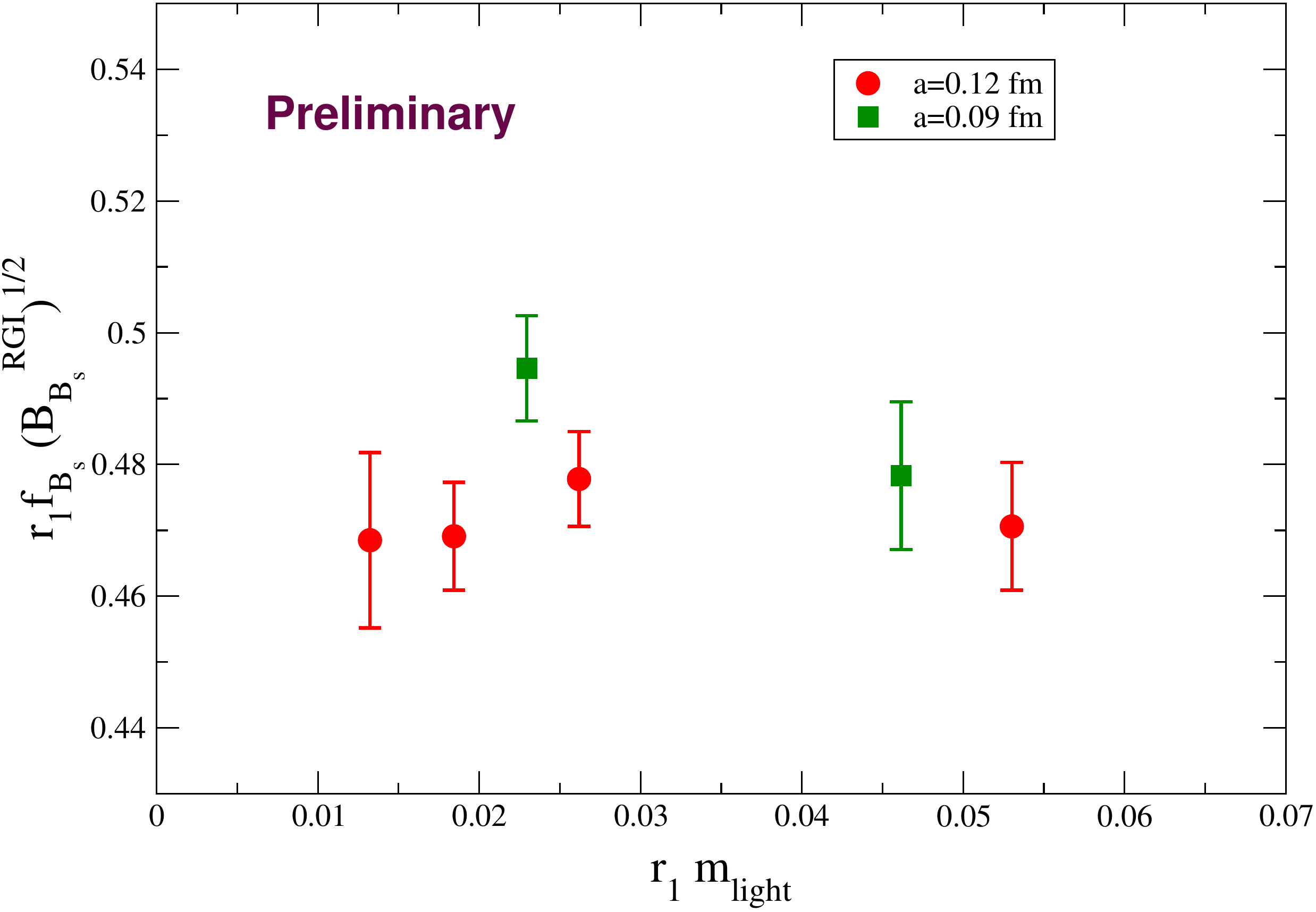}\hfill
    (b)\includegraphics[width=0.45\textwidth]{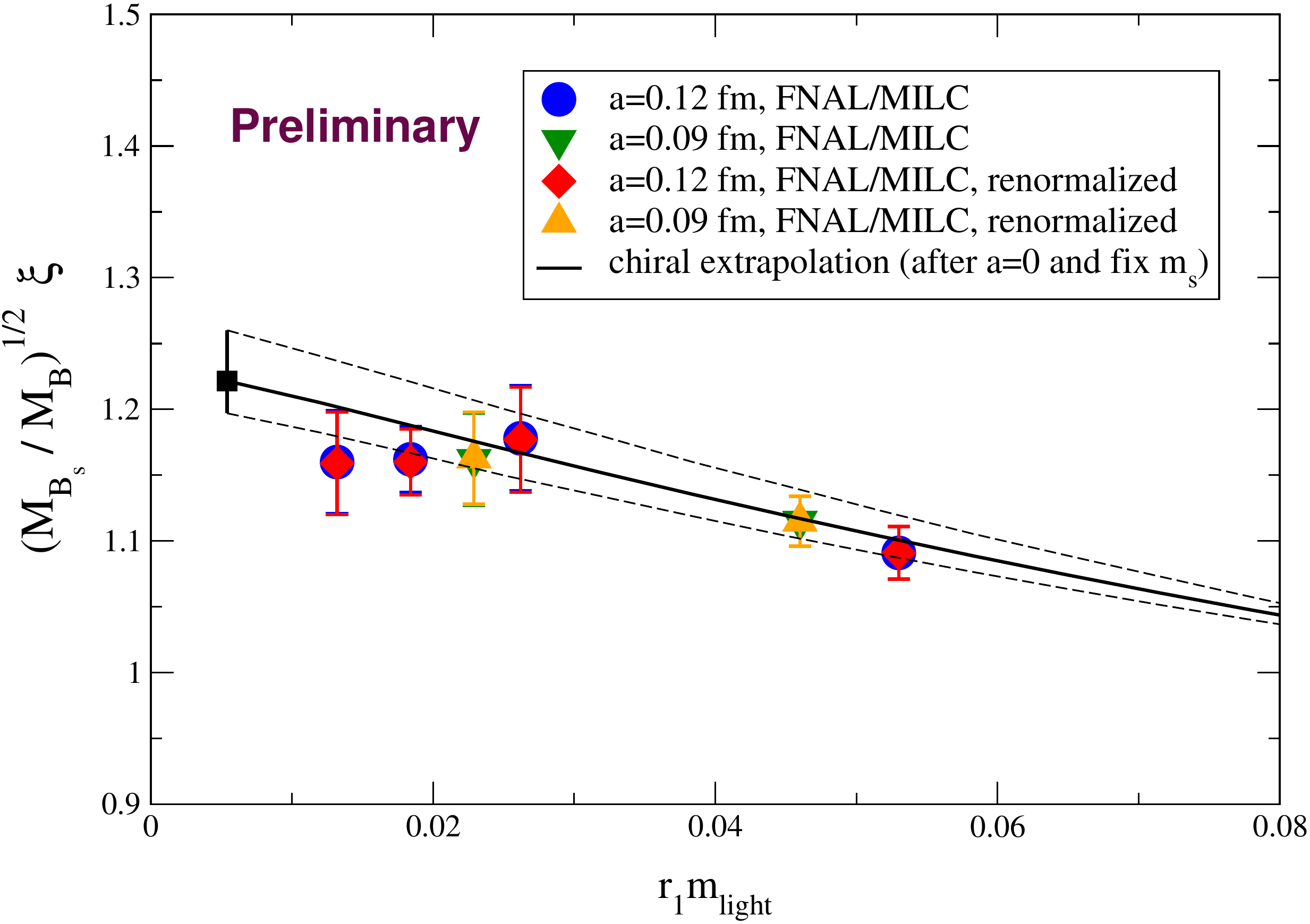}
    \caption[fig:xpt]{Light-quark-mass dependence of 
    $f_{B_s}B_{B_s}^{1/2}$ and $\xi$.
    The curve in the right plot is a fit to all partially-quenched 
    data, not just the shown unitary data.}
    \label{fig:xpt}
\end{figure}

In the phenomenology of $B$-$\bar{B}$ mixing it is conventional to 
write the matrix element as
\begin{equation}
    \langle B^0_q|\bar{q}_L\gamma_\mu b\bar{q}_L\gamma^\mu b
        |\bar{B}^0_q\rangle =
        \case{2}{3}f_{B_q}^2M^2_{B_q}B_{B_q}.
\end{equation}
Neglecting $Z-1$ and $\delta Z$ in Eq.~(\ref{eq:Q}) one sees that
$\beta_q=f_{B_q}\sqrt{M_{B_q}B_{B_q}}$.
Of special importance is
\begin{equation}
    \xi = f_{B_s}B_{B_s}^{1/2}/f_{B_d}B_{B_d}^{1/2} =
        (M_{B_d}/M_{B_s})^{1/2}(\beta_s/\beta_d),
\end{equation}
where, again, the right-most expression neglects $Z-1$ and $\delta Z$.
We use the experimentally measured meson masses and our chirally 
extrapolated $\beta_s$ and $\beta_d/\beta_s$ to obtain
$f_{B_s}B_{B_s}^{1/2}$ and $\xi$.
The light-quark-mass dependence is shown in Fig.~\ref{fig:xpt}.
\begin{table}[bp]
    \centering
    \begin{tabular}{lrrr}
        \hline\hline
        Source & $\beta_s$ & $\beta_d$ & \multicolumn{1}{c}{~~$\xi$} \\
        \hline
        Statistics    & 2.7 & 4.0 & 3.1 \\
        Scale ($r_1$) & 3.0 & 3.1 & 0.2 \\
        Sea and valence quark masses & 0.3 & 0.5 & 0.7  \\
        $b$-quark hopping parameter & $\le0.5$ & $\le0.1$ & $\le0.1$ \\
        $\chi$PT + light-quark discretization & 0.4 & 2.5 & 2.8 \\
        $g_{B^*B\pi}$ & 0.3 & 0.6 & 0.3 \\
        Heavy-quark discretization & 2 & 2 & 0.2 \\
        Matching (perturbation theory) & $\sim4$ & $\sim4$ & $\le0.5$ \\
        Finite volume & $\le0.5$ & $\le0.5$ & $\le0.1$ \\
        \hline
        Total & 6.1 & 7.3 & 4.3 \\
        \hline\hline
    \end{tabular}
    \caption[tbl:budget]{Preliminary error budget.
    Entries in percent.}
    \label{tbl:budget}
\end{table}
Further plots can be found in Ref.~\cite{ToddEvans:2007yq}.

A preliminary, but comprehensive, error budget is given in 
Table~\ref{tbl:budget}.
The $B^*$-$B$-$\pi$ coupling $g_{B^*B\pi}$ enters the expressions for the chiral 
extrapolation.
The data are not precise enough to determine $g_{B^*B\pi}$, so it must 
be set with a prior distribution in the chiral fits.
A range that encompasses phenomenological and quenched lattice estimates 
is $g_{B^*B\pi}=0.35\pm0.14$.
The error in Table~\ref{tbl:budget} corresponds to this range, while 
the prior width in the fits is~$\pm0.28$.

Until the perturbation theory has been checked, we prefer not to report 
a value for $f_{B_s}B_{B_s}^{1/2}$.
The matching corrections nearly cancel in the ratio $\beta_q/\beta_s$; 
the results with and without $Z-1$ and $\delta Z$ are nearly the same,
as shown in Fig.~\ref{fig:xpt}b.
With the error budget discussed above we find
\begin{equation}
    \xi = 1.205\pm 0.037_{\rm stat}\pm 0.034_{\rm syst},
\end{equation}
unchanged since \emph{Lattice~2008}~\cite{Evans:2008}.

\section{Future Prospects}
\label{sec:future}

When the perturbative matching has been completely checked, we will 
be in a position to present final results.
We can also compare different strategies, in particular, whether the 
perturbative expansion seems to work better for $\rho_{(S)}$ or~$Z_{(S)}$
(cf.\ Eq.~(\ref{eq:rho})).

In the longer term, we plan to obtain results for 4-quark operators 
that enter beyond the Standard Model.
Furthermore, the MILC ensembles now not only have much higher 
statistics than the current project at $a=0.12$ and $0.09$~fm, but also 
extend to smaller lattice spacings, $a=0.06$ and $0.045$~fm.
New runs with higher statistics and five lattice spacings
(also 0.15~fm) are underway.

\end{document}